\documentclass[11pt]{article}
\usepackage[a4paper,margin=30mm]{geometry}
\usepackage[T1]{fontenc}
\usepackage[utf8]{inputenc}
\usepackage{lmodern}
\usepackage{microtype}
\usepackage{amsmath,amssymb,amsthm,mathtools}
\usepackage{enumitem}
\usepackage{float}
\usepackage[hidelinks]{hyperref}

\floatstyle{ruled}
\newfloat{algorithm}{tbp}{loa}
\floatname{algorithm}{Algorithm}
\newenvironment{algorithmsteps}
 {\begin{enumerate}[label=\arabic*:,leftmargin=2.8em,itemsep=0pt,
                    parsep=0pt,topsep=2pt]}
 {\end{enumerate}}

\newtheorem{theorem}{Theorem}
\newtheorem{lemma}[theorem]{Lemma}

\theoremstyle{definition}

\theoremstyle{remark}

\newtheorem*{remark*}{Remark}

\newcommand{\SM}{\operatorname{SM}}
\newcommand{\SMI}{\operatorname{SMI}}

\title{A Single-Exponential Time Algorithm for the\\Maximum-Cardinality Weakly Stable Matching Problem\\with Ties and Incomplete Lists}
\author{Shuichi Miyazaki\thanks{Graduate School of Information Science, University of Hyogo. \texttt{shuichi@sis.u-hyogo.ac.jp}} \ and Suguru Tamaki\thanks{Graduate School of Information Science, University of Hyogo. \texttt{tamak@sis.u-hyogo.ac.jp}}}
\date{}

\begin{document}
\maketitle

\begin{abstract}
We study Stable Marriage with Ties and Incomplete lists (SMTI) under weak stability.
For an instance with $n$ agents on each side, we give an exact algorithm that
returns a maximum-cardinality weakly stable matching in $O^*(22.74^n)$ time.
To the best of our knowledge, this is the first exact algorithm for this problem running in single-exponential time.
\end{abstract}

\section{Introduction}

The stable marriage problem (SM), introduced by Gale and Shapley
\cite{GaleShapley1962}, is a fundamental problem of matching under preferences.
There are $n$ men and $n$ women, and each person ranks all members of the
opposite side in strict order of preference.
The task is to pair men and women so that no man and woman prefer each other
to their assigned partners. Such a pairing is called a stable matching.
Gale and Shapley showed that a stable matching always exists and can be found
in $O(n^2)$ time.
The problem and its extensions have practical applications, including the
assignment of medical residents to hospitals, and have been extensively
studied in algorithm theory because of their rich combinatorial structure
and diverse computational questions
\cite{GusfieldIrving1989,Manlove2013}.

Two natural extensions allow a person to be indifferent among several
partners and to regard some partners as unacceptable. These are represented
by ties and incomplete preference lists, respectively, and the resulting
problem is called Stable Marriage with Ties and Incomplete lists (SMTI).
In this paper, we consider weak stability: a mutually acceptable pair blocks
a matching only when each person is unmatched or strictly prefers the other
to the assigned partner.
Although a weakly stable matching can be found in polynomial time by
breaking ties arbitrarily and applying the Gale--Shapley algorithm,
its cardinality may depend on how the ties are broken.
It is therefore natural to seek a weakly stable matching of maximum
cardinality. This optimization problem, called \textsc{Max-SMTI}, is
NP-hard \cite{Manlove1999,IwamaEtAl1999,ManloveEtAl2002}.
Its approximability has been extensively studied, with results on both
approximation algorithms \cite{Kiraly2013,McDermid2009,Paluch2014} and hardness of approximation
\cite{DudyczEtAl2022}.

In this paper, we study exact algorithms for \textsc{Max-SMTI}.
For an instance with $n$ agents on each side, a straightforward approach
examines all $n!$ bijections between the two sides, deletes unacceptable
pairs, and checks the resulting matchings for weak stability.
This gives an $O^*(n!)=2^{\Theta(n\log n)}$-time algorithm,
where $O^*$ suppresses factors polynomial in $n$.
A natural question is whether \textsc{Max-SMTI} can be solved in single-exponential time, that is, in $O^*(c^n)$ time for some constant $c$.

\bigskip
\noindent
{\bf Our results.} 
We answer this question affirmatively by giving an exact algorithm with
running time $O^*(22.74^n)$.
To the best of our knowledge, this is the first exact algorithm for \textsc{Max-SMTI} running in single-exponential time.
Our main result, which follows from Lemmas~\ref{lem:outer-correct} and~\ref{lem:outer-runtime}, is stated as follows.

\begin{theorem}\label{thm:main}
A maximum-cardinality weakly stable matching in an SMTI instance with $n$
agents on each side can be found in $O^*(22.74^n)$ time. 
\end{theorem}

\bigskip
\noindent
{\bf Related work.} 
Irving~\cite{Irving1994} studied stable marriage with ties and introduced
weak, strong, and super-stabilities.
Manlove~\cite{Manlove1999} and Iwama et al.~\cite{IwamaEtAl1999} established the NP-hardness of maximizing the
cardinality of a weakly stable matching when ties and incomplete lists are
allowed; these results were further extended to restricted cases by
Manlove et al.~\cite{ManloveEtAl2002}.
Polynomial-time algorithms for strong stability and super-stability were
given by Irving~\cite{Irving1994} and extended to incomplete lists by
Manlove~\cite{Manlove1999}.

Approximation algorithms for \textsc{Max-SMTI} have achieved progressively
better guarantees~\cite{IwamaEtAl2007,Kiraly2008}, culminating in a
$1.5$-approximation algorithm~\cite{McDermid2009} and subsequent simpler
and faster algorithms with the same guarantee~\cite{Kiraly2013,Paluch2014}.
On the other hand, hardness results have clarified the limits of
polynomial-time approximation under various complexity
assumptions~\cite{Yanagisawa2007}.
In particular, the approximation ratio of $1.5$ is tight under the strong
Unique Games Conjecture~\cite{DudyczEtAl2022}.

Parameterized studies of SMTI have identified tractable cases and hardness
barriers using parameters that describe ties, solution size, and the
structure of the acceptability
graph~\cite{MarxSchlotter2010,AdilEtAl2018,BredereckEtAl2022}.
This line of research includes fixed-parameter algorithms and kernelization,
and has recently been extended to FPT approximation
schemes~\cite{ChenEtAl2025}.
These approaches complement our focus on the running time of exact
algorithms as a function of the number of agents.

A related algorithmic idea appears in Kavitha's single-exponential time algorithm
for the popular roommates problem with
strict preferences~\cite{Kavitha2019}. Her algorithm enumerates stable matchings in auxiliary
bipartite instances and uses an exponential bound on their number to
analyze the running time.
Our algorithm also combines stable-matching enumeration with a counting
bound, applying this approach to \textsc{Max-SMTI} through auxiliary
Hospitals/Residents instances. We use the bound of Palmer and
P\'alv\"olgyi~\cite{PalmerPalvolgyi2022} to obtain our explicit running time.

\section{Basic definitions}

Let $G=(U\cup V,E)$ be a finite bipartite graph with
$U=\{m_1,\ldots,m_n\}$ and $V=\{w_1,\ldots,w_n\}$.
The vertices of $U$ are regarded as men and the vertices of $V$ are regarded as women.
We usually use $m$ and $w$ to denote a man and a woman, respectively.
An edge $(m, w)\in E$ means that $m$ and $w$ regard each other as acceptable partners.  
A pair not in $E$ is \emph{unacceptable}.

A \emph{matching} is a set of edges no two of which share an endpoint.  If an
edge $(m, w)$ belongs to a matching $M$, then $m$ and $w$ are matched to each other and 
we write $M(m)=w$ and $M(w)=m$.  If no edge of $M$ contains $x$, then $x$ is
{\em unmatched}.  The cardinality $|M|$ is the number
of edges in $M$.  A matching between two sets of the same size is
\emph{perfect} if every vertex is matched.

For a vertex $x$, its acceptable neighbors are partitioned into nonempty \emph{tie classes}, $C_x^1, C_x^2, \ldots, C_x^{q_x}$.
A {\em preference list} of $x$ is an ordered list of these classes: $C_x^1, C_x^2, \ldots, C_x^{q_x}$.
 If $x$ has no acceptable neighbors, its preference list is empty and $q_x=0$.  
The vertex $x$ strictly prefers every member of $C_x^i$ to every member of
$C_x^j$ when $i<j$, and is indifferent between two members of the same class.
Every acceptable partner is strictly preferred to being unmatched.  If $y\in C_x^i$, the {\em rank} of $y$ in $x$'s preference list is defined to be $i$ and is written as $p_x(y)=i$.
Consequently, smaller rank means strictly better, while equal rank means a
tie. 

Let $M$ be a matching.  An acceptable pair $(m, w)\in E\setminus M$ is a
\emph{blocking pair} for $M$ (or $(m, w)$ \emph{blocks} $M$) if both of the following conditions hold:
(i) $m$ is unmatched, or $m$ strictly prefers $w$ to $M(m)$;
(ii) $w$ is unmatched, or $w$ strictly prefers $m$ to $M(w)$.
A matching is \emph{weakly stable} if it has no blocking pair.
Given the graph $G$ together with preference lists of all vertices, the problem \textsc{Max-SMTI} (Maximum Stable Marriage with
Ties and Incomplete lists) asks for a weakly stable matching of maximum
cardinality.
If $E=\varnothing$, the empty matching is optimal.
Henceforth, we assume $E\ne\varnothing$ to remove this trivial case.

The restriction of SMTI to instances without ties, i.e., each tie class contains only one agent, is called
\emph{Stable Marriage with Incomplete lists} (SMI).
If, in addition, the preference lists are complete, i.e., $G$ is a complete bipartite graph, the problem is called
\emph{Stable Marriage} (SM).

The following \emph{Hospitals/Residents} (\emph{HR}) problem,
which we use as an auxiliary construction in our algorithm,
is a many-to-one extension of SMI.
An HR instance consists of a set of residents and a set of hospitals.
We use $h$ and $r$ to denote a hospital and a resident, respectively.
Each hospital has a nonnegative integer, called a \emph{capacity}, which specifies the maximum number of residents it can receive.
Preference lists may be incomplete but must have no ties: Every
resident strictly orders his/her acceptable hospitals,
and every hospital strictly orders its acceptable residents.
A \emph{matching} is an assignment of residents to hospitals such that every assigned pair is mutually acceptable, each resident is assigned to at most one hospital, and each hospital receives at most its capacity of residents.
If resident $r$ is assigned to a hospital in $M$, we denote
that hospital by $M(r)$. We denote by $M(h)$ the set of
residents assigned to hospital $h$ in $M$.

A hospital--resident pair $(h,r)$ is a blocking pair if $h$ and $r$ find each
other acceptable, $r$ is unassigned or prefers $h$ to the currently assigned hospital,
and either (i) $h$ has fewer assigned residents than its capacity, or
(ii) $h$ prefers $r$ to at least one currently assigned resident.
A matching is stable if it has no blocking pair.  It is
\emph{capacity-filling} if every hospital receives a number of residents equal to its capacity.

\section{Description of the algorithm}\label{sec:fast}

\subsection{Basic idea}

Fix $A\subseteq U$ and $B\subseteq V$ with $|A|=|B|=s\ge1$.
Consider the problem of determining whether there exists a weakly stable matching whose matched vertices are exactly $A\cup B$, i.e., every vertex of $A$ must be matched to a vertex of $B$, and every vertex outside $A\cup B$ must remain unmatched.  
Note that stability is defined in the original instance, including vertices outside $A\cup B$; it is not merely stability within the subinstance on $A\cup B$.
Let \textnormal{\textsc{SolveFixedSets}}$(A,B)$ be a subroutine which returns such a weakly stable matching if one exists, and returns \textsc{failure} otherwise.

Our algorithm, described in Algorithm~\ref{alg:outer}, examines $s=n,n-1,\ldots,1$ in this order.
For each $s$ and every pair $A\subseteq U$, $B\subseteq V$ with $|A|=|B|=s$,
it calls \textnormal{\textsc{SolveFixedSets}}$(A,B)$.
It is not hard to see that the first successful call returns an optimal solution, which is formally stated in Lemma~\ref{lem:outer-correct}.

\begin{algorithm}[H]
\caption{Outer algorithm for \textsc{Max-SMTI}}\label{alg:outer}
\small
\noindent \textbf{Input:} an SMTI instance $G=(U\cup V,E)$ with $|U|=|V|=n$ and $E\ne\varnothing$.\par
\begin{algorithmsteps}
\item \textbf{For} $s=n,n-1,\ldots,1$, in this order, \textbf{do}:
\item \quad \textbf{For each} $A\subseteq U$, $B\subseteq V$ with $|A|=|B|=s$,
      \textbf{do}:
\item \qquad Set $M\gets\textnormal{\textsc{SolveFixedSets}}(A,B)$.
\item \qquad \textbf{If} $M\ne\textsc{failure}$, \textbf{return} $M$.
\item \quad \textbf{End for}.
\item \textbf{End for}.
\item \textbf{Return} \textsc{failure}.
\end{algorithmsteps}
\end{algorithm}

For a fixed size $s$, there are $\binom ns$ choices of $A$ and the same number of choices of $B$, so there are $\binom ns^2$ pairs $A,B$.
The total number of possible pairs is then
\[
 \sum_{s=1}^n\binom ns^2=\binom{2n}{n}-1\le4^n.
\]
Thus, if \textnormal{\textsc{SolveFixedSets}}$(A,B)$ can be implemented to run in single-exponential time
in $n$, the whole algorithm also runs in single-exponential time.

The pseudo-code of \textnormal{\textsc{SolveFixedSets}}$(A,B)$ we propose in this paper is given in Algorithm~\ref{alg:fixed}.
In the remainder of this section, we explain its behavior.

\begin{algorithm}[H]
\caption{The subroutine \textsc{SolveFixedSets}$(A,B)$}\label{alg:fixed}
\small
\noindent \textbf{Input:} sets $A\subseteq U$ and $B\subseteq V$ with
      $|A|=|B|=s\ge1$.\par
\begin{algorithmsteps}
\item \textbf{If} there is an edge $(m, w)$ such that $m \in U\setminus A$ and $w \in V\setminus B$,
      \textbf{return} \textsc{failure}.
\item Construct $F_{A,B}$ and the sets $D_m$ for all $m\in A$.
\item \textbf{If} $D_m=\varnothing$ for some $m\in A$,
      \textbf{return} \textsc{failure}.
\item Fix the tie-breaking orders $\prec_w$ and $\prec_m$, as specified in Sections~\ref{sec:graph}
      and~\ref{sec:generator}.
\item Construct an HR instance $J$ with hospital capacities left unspecified,
      and let $H$ be its hospital set.
\item \textbf{For each} capacity-vector
      $c=(c_{h_i})_{h_i\in H}\in\mathbb Z_{\ge0}^{s+1}$ with $\sum_{h_i\in H}c_{h_i}=s$,
      \textbf{do}:
\item \quad Obtain the HR instance $J_c$ from $J$ by setting the capacity of each
      hospital $h_i\in H$ to $c_{h_i}$.
\item \quad \textbf{For each} stable matching $\widehat M$ of
      $J_c$, \textbf{do}:
\item \qquad \textbf{If} $\widehat M$ is not capacity-filling,
      \textbf{continue} with the next stable matching.
\item \qquad Construct a selector $X$ from $\widehat M$ by defining $X(w_j)=m_i$ iff $\widehat M(r_j)=h_i$ for each $j$.
\item \qquad Compute $I_m(X)$ for every $m\in A$.
\item \qquad \textbf{If} $I_m(X)=\varnothing$ for some $m\in A$,
      \textbf{continue} with the next stable matching.
\item \qquad Construct $H_X$. \textbf{If} $H_X$ has a perfect matching
      $M$, \textbf{return} $M$.
\item \quad \textbf{End for}.
\item \textbf{End for}.
\item \textbf{Return} \textsc{failure}.
\end{algorithmsteps}
\end{algorithm}

\subsection{Preliminary checks}\label{sec:preprocess}
In this subsection, we explain lines 1--3 of Algorithm~\ref{alg:fixed}.

For given $A$ and $B$, if there is an acceptable edge $(m, w)$ such that $m \in U\setminus A$ and $w \in V\setminus B$, $m$ and $w$ would form a blocking pair.  We may therefore reject this choice of $A,B$ (line 1).  Henceforth assume that no such edge exists.

Let $E_{A,B}=E\cap(A\times B)$.
We define the set of {\em usable edges} $F_{A,B} \subseteq E_{A,B}$ as follows: 
$(m, w) \in F_{A,B}$ if and only if $p_m(w) \le p_m(w')$ for every $w'\in V\setminus B$ with $(m, w')\in E$, and $p_w(m) \le p_w(m')$ for every $m'\in U\setminus A$ with $(m', w)\in E$.
In other words, an edge $(m, w)\in E_{A,B}$ is usable if neither endpoint would
strictly prefer an acceptable vertex that is required to remain unmatched
to the other endpoint.
Therefore, any weakly stable matching whose matched vertices are exactly $A\cup B$ can use only edges of $F_{A,B}$.

For $m\in A$, let $D_m=\{p_m(w):(m, w)\in F_{A,B}\}$, which contains the ranks of the other endpoints of the usable edges for $m$. 
(Note that ranks always refer to the
original preference lists; they are not renumbered after unusable edges are
removed.) If $D_m=\varnothing$ for some $m$, no matching can cover $m$ using only usable
edges, so we may reject this choice of $A,B$ (line 3).
Henceforth we assume that no $D_m$ is empty.

\subsection{Rank-vectors}\label{sec:graph}
We will resume the explanation of the rest of Algorithm~\ref{alg:fixed} in the next subsection; here, we explain the ideas behind that part.

For each man $m \in A$, choose an integer $t_m\in D_m$.
Define a vector $t=(t_m)_{m\in A}$ and call it a \emph{rank-vector}. 
A rank-vector $t$ is used to restrict potential partners of $m$ to be women $w$ such that $p_m(w)=t_m$, that is, women whose rank is exactly $t_m$ in $m$'s preference list.

A rank-vector $t$ then gives a condition for potential partners of women. For each woman $w \in B$, define $Q_t(w)=\{m\in A:(m, w)\in E_{A,B},\ p_m(w)<t_m\}$ as the set of men $m$ who strictly prefer $w$ to a partner of rank $t_m$.
For each $w\in B$, break ties in her preference list arbitrarily to obtain
a strict linear order $\prec_w$ on her neighbors in $E_{A,B}$ (line 4).  
This order preserves every original strict
comparison: $p_w(m)<p_w(m')$ implies $m\prec_w m'$.
(The aim of this tie-breaking is to select a unique most-preferred man from $Q_t(w)$ when
it is nonempty. Note that the original ranks are unchanged, and stability is evaluated
with respect to the original preferences.)

For convenience, we introduce a \emph{dummy man} $m_0\notin U$, and set
$p_w(m_0)=+\infty$ for every $w\in B$.
We call a function $X:B\longrightarrow A\cup\{m_0\}$ a \emph{selector}
if, for each woman $w\in B$, it selects a man in $A$ acceptable to $w$
or the dummy man $m_0$.
For a rank-vector $t$, define its corresponding selector $X_t$ as
\[
 X_t(w)=
 \begin{cases}
  \text{the first man in }Q_t(w)\text{ in }\prec_w,
       &\text{if } Q_t(w)\ne\varnothing,\\
  m_0,& \text{if } Q_t(w)=\varnothing.
 \end{cases}
\]
Thus, when $Q_t(w)$ is nonempty, $X_t(w)$ is one of its most-preferred
men in $w$'s original preference list, selected uniquely by $\prec_w$.
In particular, $p_w(X_t(w))\le p_w(m)$ for every $m\in Q_t(w)$.
If $Q_t(w)$ is empty, $p_w(X_t(w))=+\infty$.

Now, for a rank-vector $t$, let $H_t=(A \cup B, F_t)$ be the bipartite graph where
$F_t=\{(m, w)\in F_{A,B}:p_m(w)=t_m \text{ and }p_w(m)\le p_w(X_t(w))\}$, and let $M^{*}$ be a perfect matching of $H_t$ if any.
If a man $m\in A$ strictly prefers a woman $w\in B$ to $M^{*}(m)$,
then $p_m(w)<t_m$, so $m\in Q_t(w)$.
Since $w$ is matched by an edge in $F_t$,
$p_w(M^{*}(w))\le p_w(X_t(w))\le p_w(m)$ holds,
so $w$ does not strictly prefer $m$ to her partner.
Thus no pair in $E_{A,B}$ blocks $M^{*}$.
Pairs involving vertices outside $A\cup B$ cannot block $M^{*}$ either,
by the preliminary check in line 1 and the definition of usable edges
in Section~\ref{sec:preprocess}.
Hence $M^{*}$ is a weakly stable matching that matches exactly $A\cup B$.

Conversely, let $M^*$ be a weakly stable matching whose matched vertices are
exactly $A\cup B$. 
Define a rank-vector $t^*$ by setting $t^*_m=p_m(M^*(m))$ for each $m\in A$.
Then since $M^*$ is stable, every edge of $M^*$ is usable, so $t^*_m\in D_m$ for each $m$.
Let $X_{t^*}$ be the selector corresponding to $t^*$.
If $X_{t^*}(w)=m'\in A$, then $m'\in Q_{t^*}(w)$, and weak stability implies
$p_w(M^*(w))\le p_w(m')=p_w(X_{t^*}(w))$.
If $X_{t^*}(w)=m_0$, the same inequality holds by the convention
$p_w(m_0)=+\infty$.  Thus every edge of $M^*$ is contained in $F_{t^*}$ and hence $M^*$ is a perfect matching of $H_{t^*}$.

The above observations, which will be formally established in Lemma~\ref{lem:sound}, suggest that the following procedure satisfies the requirement for \textnormal{\textsc{SolveFixedSets}}$(A,B)$: For every rank-vector $t$, construct its corresponding selector $X_t$, construct the graph $H_t$, and check whether $H_t$ has a perfect matching.
However, the number of rank-vectors can be as large as $s^s$, so this
enumeration does not provide a single-exponential running-time guarantee.
Our strategy is not to enumerate rank-vectors but to enumerate selectors instead.
We call a selector $X$ \emph{valid} if it is $X_t$ for some rank-vector $t$.
For a selector $X$, let $T(X)$ be the set of rank-vectors
corresponding to $X$.
Thus $X$ is valid if and only if $T(X)$ is nonempty.
The following facts, proved below, allow us to speed up the algorithm:
(i) the number of valid selectors is single-exponential in $s$, and 
we can enumerate all of them in time single-exponential in $s$ 
by enumerating stable matchings of a family of auxiliary HR instances; and
(ii) for each valid selector $X$, the set $T(X)$ of corresponding rank-vectors
is a direct product of sets of possible ranks, one for each man $m\in A$.
This structure allows us to construct, in polynomial time, a single bipartite
graph $H_X$ that has a perfect matching if and only if $H_t$ has a perfect
matching for some rank-vector $t\in T(X)$.
Thus, we can handle all the rank-vectors in $T(X)$ using a single graph $H_X$, instead of constructing and testing $H_t$ for each $t$ separately.

\subsection{Enumerating valid selectors efficiently}\label{sec:generator}

In this subsection, we explain how the above-mentioned ideas are implemented in Algorithm~\ref{alg:fixed}.
We first construct an auxiliary HR instance $J$ with hospital capacities
left unspecified (line 5).
In each iteration of the for-loop in lines 6--15, we choose a
\emph{capacity-vector} $c$ (line 6) and obtain the HR
instance $J_c$ by assigning these capacities to the hospitals of $J$ (line 7).
Each iteration of the inner for-loop in lines 8--14 takes a stable
matching $\widehat M$ of $J_c$ (line 8).
If $\widehat M$ is not capacity-filling, we skip this iteration (line 9).
Otherwise, we obtain a selector $X$ from $\widehat M$ (line 10), as detailed below.
We test whether $X$ is valid in lines 11--12.
For each selector that passes this test, we construct the bipartite graph
$H_X$ and test whether it has a perfect matching (line 13).

We now describe the construction of $J$ in line 5.
For each woman $w_j\in B$, create a resident $r_j$.
For each man $m_i\in A \cup \{ m_0 \}$, create a hospital $h_i$, and let $H$ denote the set of hospitals.
We call $h_0$ the \emph{dummy hospital}, and all other hospitals \emph{real hospitals}.
Thus, $J$ has $s$ residents and $s+1$ hospitals.
Recall that the women's tie-breaking orders $\prec_w$ were defined
in Section~\ref{sec:graph}. For each man $m\in A$, break ties in his
preference list similarly to obtain a strict order $\prec_m$ (line 4).
We use these same orders throughout this call to \textsc{SolveFixedSets}.
To construct resident $r_j$'s preference list in $J$, start with
woman $w_j$'s original preference list, delete men not in $A$,
break ties according to $\prec_{w_j}$, replace each
remaining man $m_i$ by hospital $h_i$, and
append $h_0$ to the end of the list.
To construct real hospital $h_i$'s preference list, start with man $m_i$'s
original preference list, delete women not in $B$, break ties according to
$\prec_{m_i}$, and replace each remaining woman $w_j$ by resident $r_j$.
The preference list of $h_0$ contains all residents $r_j$ in an arbitrary but fixed strict order.

Once $J$ has been constructed, we enumerate its possible hospital capacities.
A \emph{capacity-vector} is a vector
$c=(c_{h_i})_{h_i\in H}\in\mathbb Z_{\ge0}^{s+1}$, where $c_{h_i}$
specifies the capacity of hospital $h_i$.
The loop in line 6 considers all such vectors with
$\sum_{h_i\in H}c_{h_i}=s$.
For each $c$, let $J_c$ be the HR instance obtained from $J$ by setting
the capacity of each hospital $h_i\in H$ to $c_{h_i}$ (line 7).
Note that the agents and preference lists are fixed by $J$; only the hospital
capacities vary with $c$.
Note also that hospitals with capacity 0 are retained in $J_c$, though they do not play any role.

To enumerate stable matchings of $J_c$ in line 8, apply the standard
cloning construction \cite[p.~38]{GusfieldIrving1989} to obtain an SMI instance $K_c$.
Since $J_c$ has $s$ residents and total capacity $s$, $K_c$
has $s$ agents on each side.  The construction gives a one-to-one
correspondence between the stable matchings of $J_c$ and $K_c$.
We enumerate the stable matchings of $K_c$ using the
rotation-based enumeration method \cite{Gusfield1987,GusfieldIrving1989},
and convert each into the corresponding stable matching $\widehat M$
of $J_c$.
If $\widehat M$ is not capacity-filling, discard it (line 9);
otherwise define a selector $X:B\to A\cup\{m_0\}$ as follows (line 10).
For each $j$, if resident $r_j$ is assigned to hospital $h_i$ in
$\widehat M$, set $X(w_j)=m_i$.
By this correspondence and the capacity-filling property of $\widehat M$,
for each hospital $h_i\in H$, we have $| X^{-1}(m_i) | = |\widehat M(h_i) | =c_{h_i}$.
Thus, the capacity of hospital $h_i$ is also the number of times that
the corresponding man $m_i$ is selected by $X$.

We then test whether $X$ is valid as follows (lines 11--12). Define 
\begin{align*}
 L_m(X)
  &=\max\bigl(\{p_m(w):w\in B,\ X(w)=m\}\cup\{-\infty\}\bigr),\\
 R_m(X)
  &=\min\Bigl(
     \{p_m(w):(m, w)\in E_{A,B},\ X(w)=m_0\}\\
  &\hspace{18mm}{}\cup
     \{p_m(w):(m, w)\in E_{A,B},\ X(w)\in A,\ m\prec_w X(w)\}
     \cup\{+\infty\}\Bigr),
\end{align*}
and let $I_m(X)=\{d\in D_m:L_m(X)<d\le R_m(X)\}$ for each $m\in A$.
If $I_m(X)=\varnothing$ for some $m\in A$, we decide that $X$ is not valid (line 12).
Otherwise, we decide that $X$ is valid.
The correctness of this validity test is proved in Lemma~\ref{lem:meaning}(i).
If $X$ is valid, we construct the bipartite graph $H_X=(A\cup B,E_X)$ where $E_X=\{(m,w)\in F_{A,B}:L_m(X)<p_m(w)\le R_m(X),\, p_w(m)\le p_w(X(w))\}$, and test whether it has a perfect matching (line 13).

\begin{remark*}
For a fixed capacity-vector $c$, the Rural Hospitals Theorem \cite{Roth1986}
implies that each hospital receives the same number of residents in every
stable matching of $J_c$.
Hence, if a stable matching fails the capacity-filling check in line 9,
no stable matching of $J_c$ is capacity-filling.
We may therefore discard this $J_c$ and proceed to the
next capacity-vector in the outer for-loop.
For simplicity, however, this speed-up is not implemented in Algorithm~\ref{alg:fixed}.
\end{remark*}

\section{Correctness}\label{sec:fast-correct}

In this section, we first prove several lemmas and then establish the correctness
of Algorithm~\ref{alg:outer} in Lemma~\ref{lem:outer-correct} at the end of the section.
In these lemmas, the bipartite graph $H_t$ is the one defined in Section~\ref{sec:graph}.
Although it is not directly used in Algorithm~\ref{alg:fixed}, its properties and its connection to $H_X$ play important roles in proving correctness.

\begin{lemma}\label{lem:sound}
Let $A$ and $B$ be subsets of $U$ and $V$, respectively, with
$|A|=|B|\ge1$, that pass the checks in lines 1 and 3 of Algorithm~\ref{alg:fixed}.
Then the following hold.
(i) For every rank-vector $t$, every perfect matching $M$ of $H_t$
is weakly stable and matches exactly the vertices in $A\cup B$.
(ii) For every weakly stable matching $M$ whose matched vertex set is exactly
$A\cup B$, the vector $t^*$ defined by $t^*_m=p_m(M(m))$ for each $m\in A$
is a rank-vector, and $M$ is a perfect matching of $H_{t^*}$.
\end{lemma}
\begin{proof}
For (i), let $M$ be a perfect matching of $H_t$.
By the definition of $H_t$, $p_m(M(m))=t_m$ for every $m\in A$.
Let $(m, w) \in E \setminus M$.
We check whether $(m, w)$ blocks $M$.
By the check in line 1, $m \in A$ or $w \in B$ holds.
Suppose that $m \in A$ and $w \not\in B$.
Then $p_m(M(m)) \leq p_m(w)$ because $(m, M(m))$ is usable.
Hence $(m, w)$ does not block $M$.
The case of $m \not\in A$ and $w \in B$ can be argued similarly.
Finally, suppose that $m \in A$ and $w \in B$.
If $p_m(w)\ge t_m$, then $p_m(M(m))=t_m\le p_m(w)$,
so $m$ does not strictly prefer $w$ to his partner $M(m)$.  Otherwise
$m\in Q_t(w)$, and by the definition of $X_t(w)$, we have $p_w(M(w))\le p_w(X_t(w))\le p_w(m)$,
so $w$ does not strictly prefer $m$ to her partner $M(w)$.
Therefore, in either case $(m, w)$ cannot block $M$.

For (ii), let $M$ be a weakly stable matching whose matched vertex set is
exactly $A\cup B$. Since $t^*_m=p_m(M(m))$ and every edge of $M$ is usable, $t^*_m\in D_m$ for each $m\in A$,
so $t^*$ is a rank-vector. 
We show that $M$ is a perfect matching of $H_{t^*}$.
Consider the selector $X_{t^*}$ corresponding to $t^*$.
For any woman $w\in B$, if $X_{t^*}(w)=m'\in A$, then
$m'\in Q_{t^*}(w)$, so $p_{m'}(w)<t_{m'}^*=p_{m'}(M(m'))$.
Thus $m'$ strictly prefers $w$ to his partner in $M$.
Weak stability of $M$ therefore implies
$p_w(M(w))\le p_w(m')=p_w(X_{t^*}(w))$.
If $X_{t^*}(w)=m_0$, the same inequality holds because
$p_w(m_0)=+\infty$.
Since every edge of $M$ is usable and $p_m(M(m))=t_m^*$ for every
$m\in A$, these inequalities show that every edge of $M$ belongs to
$F_{t^*}$.  Hence $M$ is a perfect matching of $H_{t^*}$.
\end{proof}

Lemma~\ref{lem:meaning}(i) justifies the validity test in lines 11--12,
while (ii) characterizes the set of rank-vectors corresponding to a selector.

\begin{lemma}\label{lem:meaning}
Let $A$ and $B$ be subsets of $U$ and $V$, respectively, with
$|A|=|B|\ge1$.
Let $X:B\to A\cup\{m_0\}$ be a selector.
Then, (i) $X$ is valid if and only if $I_m(X)\ne\varnothing$
for every $m\in A$, and (ii) $T(X)=\prod_{m\in A}I_m(X)$.
\end{lemma}
\begin{proof}
Consider a woman $w\in B$ and a rank-vector $t$.
By definition, for a man $m\in A$ acceptable to $w$, $X_t(w)=m$ holds exactly when $p_m(w)<t_m$ and
$t_{m'}\le p_{m'}(w)$ for every man $m'\in A$ acceptable to $w$ such that $m'\prec_w m$.
Also, $X_t(w)=m_0$ holds exactly when $t_{m'}\le p_{m'}(w)$ for every man $m'\in A$ acceptable to $w$.
Rearranging these conditions by man, we see that, for a given selector $X$,
$X_t=X$ holds if and only if the following conditions hold for every man $m\in A$:
\begin{enumerate}
\item[(a)] for every woman $w\in B$ such that $X(w)=m$, we have $p_m(w)<t_m$;
\item[(b)] for every woman $w\in B$ such that $(m,w)\in E_{A,B}$ and
$X(w)=m_0$, we have $t_m\le p_m(w)$; and
\item[(c)] for every woman $w\in B$ such that $(m,w)\in E_{A,B}$,
$X(w)\in A$, and $m\prec_w X(w)$, we have $t_m\le p_m(w)$.
\end{enumerate}
By the definitions of $L_m(X)$ and $R_m(X)$, conditions (a)--(c) are equivalent to ``$L_m(X)<t_m\le R_m(X)$.''
Since a rank-vector satisfies $t_m\in D_m$, this is in turn
equivalent to ``$t_m\in I_m(X)$.''
Hence if $X$ is valid, there is a corresponding rank-vector $t$ and for each $m\in A$, $t_m\in I_m(X)$. Thus $I_m(X)\ne\varnothing$ for every $m\in A$.
Conversely, if $I_m(X)\ne\varnothing$ holds for every $m\in A$, we can define a rank-vector $t$ in such a way that $t_m\in I_m(X)$ for every $m\in A$. This $t$ satisfies conditions (a)--(c), so $X_t=X$ and hence $X$ is valid.
Therefore (i) holds.

For (ii), by the above argument, any choice of $t_m$ from $I_m(X)$ for each $m$ results in a rank-vector $t$ satisfying $X_t=X$.
Hence $T(X)=\prod_{m\in A}I_m(X)$.
\end{proof}

The next lemma relates perfect matchings of $H_X=(A\cup B,E_X)$ and $H_t=(A \cup B, F_t)$.

\begin{lemma}\label{lem:preserve}
Let $A$ and $B$ be subsets of $U$ and $V$, respectively, with
$|A|=|B|\ge1$.
Let $X:B\to A\cup\{m_0\}$ be a valid selector.
Then, (i) for every $t\in T(X)$, any perfect matching $M$ of $H_t$
is a perfect matching of $H_X$, and (ii) for any perfect matching $M$ of $H_X$,
there exists a rank-vector $t$ such that $M$ is a perfect matching of $H_t$.
\end{lemma}
\begin{proof}
\textup{(i)} Let $t\in T(X)$. Then $X_t=X$.
By Lemma~\ref{lem:meaning}(ii), we have
$L_m(X)<t_m\le R_m(X)$ for every $m\in A$.
Let $(m,w)$ be an arbitrary edge in $F_t$.
By the definition of $F_t$, we have $(m,w)\in F_{A,B}$, $p_m(w)=t_m$, and $p_w(m) \le p_w(X_t(w))$.
These equalities and inequalities imply $L_m(X) <p_m(w) \le R_m(X)$ and $p_w(m) \le p_w(X(w))$, which is the condition for $(m,w)\in E_X$.
Thus $F_t\subseteq E_X$.  Since $H_t$ and $H_X$ have the same vertex set,
every perfect matching of $H_t$ is a perfect matching of $H_X$.

\textup{(ii)} Let $M$ be a perfect matching of $H_X$ and let $(m,w)$ be an arbitrary edge of $M$.
Since $(m,w)\in E_X$, we have $(m,w)\in F_{A,B}$, $L_m(X)< p_m(w) \le R_m(X)$, and $p_w(m)\le p_w(X(w))$.
Since $(m,w)\in F_{A,B}$, $(m,w)$ is a usable edge, so if we define a vector $t$ by $t_m=p_m(w)$ for each $(m, w) \in M$, $t_m \in D_m$ and thus $t$ is a rank-vector.
Because $t_m\in D_m$ and $L_m(X)< t_m \le R_m(X)$, $t_m\in I_m(X)$ for every $m\in A$.
Then, by Lemma~\ref{lem:meaning}(ii), $t\in T(X)$, so we have $X_t=X$.
Therefore, we have $p_w(m)\le p_w(X_t(w))$.
This, together with $(m,w)\in F_{A,B}$ and $t_m=p_m(w)$, implies $(m,w)\in F_t$.
Thus $M$ is a perfect matching of $H_t$.
\end{proof}

The next lemma shows that every valid selector corresponds to a
capacity-filling stable matching of some $J_c$.

\begin{lemma}\label{lem:hr}
Let $A$ and $B$ be subsets of $U$ and $V$, respectively, with
$|A|=|B|\ge1$.
For a valid selector $X$, let $c$ be the capacity-vector defined by
$c_{h_i}=|X^{-1}(m_i)|$ for each hospital $h_i\in H$.
Define an assignment $M$ by assigning, for each $j$, resident $r_j$
to hospital $h_i$ if $X(w_j)=m_i$.
Then, $M$ is a capacity-filling stable matching of $J_c$.
\end{lemma}
\begin{proof}
Consider a woman $w_j\in B$, and let $m_i=X(w_j)$.
If $i=0$, then $(h_0,r_j)$ is an acceptable pair in $J_c$, since the
dummy hospital and every resident find each other acceptable.
If $i\ge1$, then $(m_i,w_j)\in E_{A,B}$ by the definition of a selector.
Hence, the corresponding pair $(h_i,r_j) \in M$ is acceptable in $J_c$.
By the construction of $M$ and the definition of $c$, every resident is assigned exactly once, and each hospital $h_i$ is assigned exactly $c_{h_i}$ residents.
Thus, $M$ is a capacity-filling matching of $J_c$.
It remains to show that $M$ is stable.

Consider an acceptable pair $(h_i,r_j)$ in $J_c$ that is not in $M$.
If $i=0$, then $r_j$ is assigned to a real hospital, which $r_j$ prefers to
$h_0$, so $(h_0,r_j)$ does not block $M$.
Suppose then that $i\ge1$.
Since $X$ is valid, there is a rank-vector $t$ such that $X_t=X$, so fix this $t$.
If $m_i\in Q_t(w_j)$, let $h_k=M(r_j)$.
By the construction of $M$, we have $X_t(w_j)=m_k$.
Then $m_k \prec_{w_j} m_i$ by the definition of $X_t(w_j)$,
so resident $r_j$ prefers $h_k$ to $h_i$ in $J_c$.
Thus, $(h_i,r_j)$ does not block $M$.
If $m_i\notin Q_t(w_j)$, then $t_{m_i}\le p_{m_i}(w_j)$.
On the other hand, every resident $r_\ell \in M(h_i)$ satisfies
$X_t(w_\ell)=m_i$, so $m_i\in Q_t(w_\ell)$ and
$p_{m_i}(w_\ell)<t_{m_i}$.
Consequently, $p_{m_i}(w_\ell)<t_{m_i}\le p_{m_i}(w_j)$ for every such $r_\ell$.
By the construction of $J_c$, hospital $h_i$ strictly prefers $r_\ell$ to $r_j$.
Since $M$ is capacity-filling, $(h_i,r_j)$ does not block $M$.
Hence $M$ is stable.
\end{proof}

\begin{lemma}\label{lem:fixed-correct}
Algorithm~\ref{alg:fixed} returns a weakly stable matching whose matched vertex set is exactly
$A\cup B$ if one exists, and returns \textsc{failure} otherwise.
\end{lemma}
\begin{proof}
First, suppose that Algorithm~\ref{alg:fixed} returns a matching $M$ in line 13.
Then, $A$ and $B$ passed the preliminary checks in lines 1 and 3.
Let $X$ be the selector used in the iteration that returns $M$.
Then $M$ is a perfect matching of $H_X$.
The selector $X$ passes the check in line 12 and is therefore valid
by Lemma~\ref{lem:meaning}(i).
Then, by Lemma~\ref{lem:preserve}(ii), there exists a rank-vector $t$
such that $M$ is a perfect matching of $H_t$.
By Lemma~\ref{lem:sound}(i), $M$ is weakly stable and matches vertex set $A\cup B$.

Conversely, suppose that there is a weakly stable matching $M$ whose
matched vertex set is exactly $A\cup B$, and let $s=|A|=|B|$.
There can be no edge between $U\setminus A$ and $V\setminus B$, since
its endpoints would block $M$; hence the check in line 1 does not reject
$A,B$.  Since $M$ is stable, every edge of $M$ is usable, so
$p_m(M(m))\in D_m$ for every $m\in A$; hence the check in line 3
does not reject $A,B$ either.
Define the rank-vector $t^*$ by $t_m^*=p_m(M(m))$ for each $m\in A$.
Then, by Lemma~\ref{lem:sound}(ii), $M$ is a perfect matching of $H_{t^*}$, and by Lemma~\ref{lem:preserve}(i), $M$ is also a perfect matching of $H_{X_{t^*}}$.

Since $X_{t^*}$ is valid, once $X_{t^*}$ is generated in line 10, this iteration is not skipped
in line 12 by Lemma~\ref{lem:meaning}(i).  
Line 13 then returns a perfect matching of $H_{X_{t^*}}$,
though not necessarily $M$.
By the first part of this proof, the returned matching has the required
properties.  The same is true if another matching is returned earlier for another valid selector.
It therefore remains to show that $X_{t^*}$ is generated in line 10 unless
a matching has already been returned.

To this end, let $c^*$ be the capacity-vector defined by
$c_{h_i}^*=|(X_{t^*})^{-1}(m_i)|$ for every hospital $h_i\in H$.
Since $\sum_{h_i\in H}c_{h_i}^*=s$, $c^*$ is enumerated in
line 6, and setting the hospital capacities of $J$ to $c^*$ in line 7
gives $J_{c^*}$.
Let $\widehat M^*$ be the assignment obtained by assigning, for each $j$,
resident $r_j$ to hospital $h_i$ if $X_{t^*}(w_j)=m_i$.
By Lemma~\ref{lem:hr}, $\widehat M^*$ is a capacity-filling stable
matching of $J_{c^*}$, so it is enumerated in line 8 and passes the
check in line 9.
By the construction of $\widehat M^*$, $X_{t^*}$ is generated from $\widehat M^*$
in line 10.
\end{proof}

\begin{lemma}\label{lem:outer-correct}
Algorithm~\ref{alg:outer} returns a maximum-cardinality weakly stable matching.
\end{lemma}
\begin{proof}
Let $M^*$ be a maximum-cardinality weakly stable matching, and let
$s^*=|M^*|\ge1$.  Let $A^*$ and $B^*$ be its matched vertex sets
on the two sides.
By Lemma~\ref{lem:fixed-correct}, every successful call to
\textsc{SolveFixedSets}$(A,B)$ returns a weakly stable matching of size $s$.
Hence no call with $s>s^*$ succeeds, whereas the call with
$A=A^*$ and $B=B^*$ succeeds if reached.
Since Algorithm~\ref{alg:outer} examines the sizes in decreasing order,
its first successful call returns a weakly stable matching of size $s^*$,
which is optimal.
\end{proof}

\section{Running time}\label{sec:runtime}
In this section, we analyze the running time of Algorithm~\ref{alg:outer}.
Since the $O^*$ notation suppresses factors polynomial in $n$,
we do not evaluate their degrees explicitly.
We first bound the running time of each call to Algorithm~\ref{alg:fixed}.

\begin{lemma}\label{lem:fixed-runtime}
For $A\subseteq U$ and $B\subseteq V$ with $|A|=|B|=s\ge1$,
Algorithm~\ref{alg:fixed} runs in $O^*(14.2^s)$ time.
\end{lemma}
\begin{proof}
We follow the steps of Algorithm~\ref{alg:fixed}.
The checks and constructions in lines 1--3, the choice of tie-breaking
orders in line 4, and the construction of $J$ in line 5 take polynomial time.

The outer for-loop in line 6 considers capacity-vectors $c$ for the
$s+1$ hospitals in $H$, each consisting of nonnegative integers
with sum $s$.
By the stars-and-bars formula, the number of such capacity-vectors is
\[
 \binom{s+(s+1)-1}{(s+1)-1}=\binom{2s}{s}.
\]
These vectors can be enumerated in polynomial time per vector.
In each iteration, setting the hospital capacities of $J$ to $c$ to
obtain $J_c$ in line 7 and applying the cloning construction to obtain
$K_c$ take polynomial time.

We next bound the number of iterations of the inner for-loop in line 8.
As described in Section~\ref{sec:generator}, this loop is implemented by
enumerating the stable matchings of the SMI instance $K_c$, which has
$s$ agents on each side.
Let $\SM(s)$ (resp.~$\SMI(s)$) be the maximum number of stable matchings in an SM (resp.~SMI) instance
with $s$ agents on each side.
It is not hard to see that $\SMI(s) \leq \SM(s)$.
Also, Palmer and P\'alv\"olgyi \cite{PalmerPalvolgyi2022} proved that there is an absolute constant $C$ such that
$\SM(s)\le C \cdot 3.55^s$ for every $s\ge1$.
Thus the inner for-loop has at most $C \cdot 3.55^s$ iterations for each capacity-vector $c$.

For SMI instances, all stable matchings can be enumerated in polynomial time per output, after polynomial-time preprocessing, using the rotation representation \cite{Gusfield1987,GusfieldIrving1989}.
Converting each output stable matching of $K_c$ into the corresponding
stable matching of $J_c$ also takes polynomial time.
Hence the total time needed to enumerate the stable matchings for each
capacity-vector $c$ is $O^*(3.55^s)$.

In each inner iteration, checking whether the matching is capacity-filling
in line 9, defining $X$ in line 10, computing $L_m(X)$, $R_m(X)$, and
$I_m(X)$ in line 11, and checking whether any $I_m(X)$ is empty in
line 12 take polynomial time.
Constructing $H_X$ and finding a perfect matching, if one exists,
in line 13 also take polynomial time.
Therefore, for fixed $A,B$ of size $s$, the running time of
Algorithm~\ref{alg:fixed} is
\[
 O^*\!\left(\binom{2s}{s}3.55^s\right) \le O^*\!\left(4^s\cdot \,3.55^s\right) =O^*(14.2^s).
\]
Here we used $\binom{2s}{s}\le2^{2s}=4^s$.
\end{proof}

Now we are ready to evaluate the running time of Algorithm~\ref{alg:outer}.

\begin{lemma}\label{lem:outer-runtime}
Algorithm~\ref{alg:outer} runs in $O^*(22.74^n)$ time.
\end{lemma}
\begin{proof}
As mentioned in Section~\ref{sec:fast}, for each $s$ there are
$\binom ns^2$ pairs $(A,B)$ with
$|A|=|B|=s$.  By Lemma~\ref{lem:fixed-runtime}, each call to
Algorithm~\ref{alg:fixed} takes $O^*(14.2^s)$ time,
so the total running time of Algorithm~\ref{alg:outer} is at most
\begin{align*}
 O^*\!\left(\sum_{s=1}^n \left( \binom ns^2 14.2^s\right) \right)
 &= O^*\!\left(
       \sum_{s=1}^n \left( \binom ns(\sqrt{14.2})^s\right)^2
       \right)\\
 &\le O^*\!\left(
       \left(\sum_{s=1}^n\binom ns(\sqrt{14.2})^s\right)^2
       \right)\\
 &=O^*\!\left(\bigl((1+\sqrt{14.2})^n-1\bigr)^2\right)\\
 &\le O^*\!\left((1+\sqrt{14.2})^{2n}\right).
\end{align*}
The second equality follows from the binomial theorem.  Since $(1+\sqrt{14.2})^2=22.736577\ldots<22.74$, the running time is $O^*(22.74^n)$.
\end{proof}

\section*{Use of generative AI}
During the development and preparation of this work, the authors used
ChatGPT (OpenAI) to discuss algorithmic ideas and proof arguments, explore
examples and counterexamples, assist with literature searches, and draft
and revise the English exposition and \LaTeX{} source.
The authors reviewed and revised the AI-assisted material and retained
control over the research and the manuscript.
The authors take full responsibility for the correctness of the
mathematical results and proofs, the accuracy of the references, and the
final content of the manuscript.

\section*{Acknowledgments}
This work was partially supported by JSPS KAKENHI Grant Numbers JP22K11909 and JP25K14996, and the joint project of Kyoto University and Toyota Motor Corporation, titled ``Advanced Mathematical Science for Mobility Society.''

\end{document}